\documentclass[a4paper,11pt]{article}

\usepackage{authblk}
\usepackage[utf8]{inputenc}
\usepackage[T1]{fontenc}
\usepackage{amsfonts} 
\usepackage{graphicx}
\usepackage{epstopdf}
\usepackage{enumitem} 
\usepackage{amsopn}   
\usepackage{xspace}   

\usepackage{amsthm}

\usepackage{amsmath,amssymb,enumitem}

\usepackage{algorithm}
\usepackage{algpseudocode}

\newtheorem{theorem}{Theorem}
\newtheorem{proposition}[theorem]{Proposition}
\newtheorem{lemma}[theorem]{Lemma}
\newtheorem{corollary}[theorem]{Corollary}
\newtheorem{definition}[theorem]{Definition}
\newtheorem{claim}[theorem]{Claim}

\DeclareMathOperator{\Probability}{\mathbb{P}}

\newcommand{\Prob}[1]{\Probability\pbrcx{#1}}

\newcommand{\R}{{\mathbb{R}}}

\newcommand{\N}{{\mathbb{N}}}

\newcommand{\pth}[1]{\ensuremath{\left(#1\right)}}
\newcommand{\pbrcx}[1]{\ensuremath{\left[#1\right]}}

\newcommand{\calG}{{\cal G}}

\newcommand{\calH}{{\cal H}}

\newcommand{\Findcenter}{{\rm{FindCenter }}}
\newcommand{\TvbgPt}{{\rm{TverbergPoint }}}

\newcommand{\e}{{\varepsilon}}

\newcommand{\calR}{{\mathcal{R}}}
\newcommand{\calF}{{\mathcal{F}}}

\newcommand{\conv}{{\rm conv}}

\newcommand*\samethanks[1][\value{footnote}]{\footnotemark[#1]}

\title{Near-Optimal Centerpoints in Polynomial Time in the Ambient Dimension\footnote{A preliminary version of this article appeared in the 
Proceedings of the 2026 Annual ACM-SIAM Symposium on Discrete Algorithms (SODA 2026).}}

\author[1]{Kunal Dutta
\thanks{ Supported by the Polish NCN grant nr. 2023/51/B/ST6/02989.
}
}

\author[2]{Karol Pisula
\samethanks[1]
}

\affil[1]{Institute of Informatics, University of Warsaw, Poland. email: kdutta@mimuw.edu.pl}
\affil[2]{Institute of Informatics, University of Warsaw, Poland. email: k.pisula@student.uw.edu.pl}

\date{}

\begin{document}
\maketitle

\begin{abstract}
An $\alpha$-centerpoint of a set of $n$ input points in $\R^d$ is a point such that any halfspace containing it also contains an $\alpha$-fraction of the input points. Since the first algorithm for centerpoints given by Clarkson, Eppstein, Miller, 
Sturtivant and Teng [IJCGA, 1996], the best quality centerpoint obtainable in polynomial time in $d$ has been 
$\Omega(d^{-2})$, while the best possible centerpoints of $\Omega(d^{-1})$ quality could only be obtained 
in time $d^{\Omega(d)}$, (Clarkson et al. [IJCGA, 1996]).
Recently Cherapanamjeri [\emph{FOCS}, 2024] gave the first polynomial time randomized algorithm for 
computing \emph{approximate} $\Omega(1/d)$-centerpoints. However it has run-time 
significantly worse than $O(d^{16})$, as well as large constants like $2^{80}$. 

We give a $\tilde{O}(d^5)$ time randomized algorithm for computing \emph{exact} $\Omega(1/(d\cdot\ln d\cdot\ln\ln d))$-centerpoints of arbitrary point-sets. This improves significantly on the long-standing $\Omega(1/d^2)$ quality bound of Clarkson et al. 
[IJCGA, 1996], (see also Har-Peled and Jones [SoCG 2019]), and is within a 
$\Omega((\ln d\ln\ln d)^{-1})$-factor of the optimum. It is significantly simpler to implement, and improves  
on the longstanding $d^{O(d)}$ running time of Clarkson et al. [\emph{IJCGA}, 1996] for $\omega(1/d^2)$-centerpoints. 

For low-dimensional point-sets, we also give the first centerpoint algorithm with complexity subexponential in the number of dimensions, 
and independent of the data bit-complexity, which achieves 
a $\omega\pth{d^{-2}}$ quality of the centerpoint,  
improving on the previous $O\pth{d^d}$ bound.  
For the Word RAM model, as well as for practical implementations, we also discuss a truncate-and-proceed approach to address bit-size complexity issues.

On the way we also get a simple algorithm to compute $O\pth{\sqrt{\frac{n}{d}}}$-Tverberg partitions, and a point in the 
intersection of the convex hulls of the Tverberg parts. Our central idea is based on a lifting 
trick of Sarkaria [\emph{Isr. J. Math.} 1991], together with sampling ideas from 
Sober\'on [\emph{Combinatorica}, 2019]. 
At the core of our algorithmic ideas, is a randomized algorithm for the \emph{Colorful 
Carath\'eodory} problem, 
which improves upon the $d^{\Theta(\log d)}$ run-time of Mulzer and Stein [SoCG, 2015].

Lastly, we also give applications of our improved bounds to problems such as the Resilient Vector Consensus problem, 
lower-bounding functions with gradient oracles and deciding the lightness of a convex body.
\end{abstract}

        \section{Introduction}
        \label{sec:intro}

Given an arbitrary set $P$ of $n$ points in $d$-dimensional space $\R^d$, a classic result of discrete geometry states that 
there exists a point of $\R^d$, such that any closed halfspace containing this point, also contains at least $\frac{n}{d+1}$ points of 
the given set. This point is called the \emph{centerpoint} of the set $P$. More generally, for a given \emph{quality} $\alpha\in [0,1]$, an $\alpha$-centerpoint  
of a point-set $P$ is a point $x \in \R^d$ such that any closed halfspace containing $x$ also contains $\alpha n$ points of $P$. 

The notion of centerpoints arose in discrete geometry and linear algebra, and since then has reappeared in other areas such as topology, 
statistics, etc.~\cite{M02,DBLP:conf/soda/Chan04}. More recently, 
it has found algorithmic applications in the analysis of large datasets in fields like data science, machine learning, and topological 
data analysis~\cite{DBLP:journals/simods/LoeraH20,DBLP:journals/dcg/LoeraHOY21}.    
In many of these applications, the centerpoint serves as a decent approximation as the ``center" of the dataset. The centerpoint is also 
closely related to several other geometric notions of centrality of datasets, such as Radon and Tverberg points, points of high Tukey depth, etc. 

As such the algorithmic question of finding the centerpoint of a given $d$-dimensional dataset, or at least an $\alpha$-centerpoint for some 
slightly suboptimal $\alpha$, is of significant interest and has received the attention of several 
researchers~\cite{DBLP:journals/ijcga/ClarksonEMST96,DBLP:conf/soda/Chan04,DBLP:conf/compgeom/AgarwalSW04,DBLP:journals/talg/Har-PeledJ21}.

Before discussing the previous works, we need to define a few models of computation that will be relevant to the subsequent discussion. In the present 
work, and as in typical in computational geometry, 
we use the Real RAM model of computation, where real numbers can be stored using a single unit of memory, and basic arithmetic operations require a single 
unit of computation time. In contrast, in the Word RAM model, each bit of an input number requires a single unit of memory. 
An algorithm is said to have a \emph{weakly polynomial} running time, if it runs in time polynomial in the size of its input in bits. It is \emph{strongly polynomial}, 
        if the running time is independent of the size of the input in bits. We shall use the $O_w(.)$-notation to denote a weakly polynomial running time, where the dependence on the size 
        of the inputs will be suppressed. We also use $\tilde{O_w}$- to indicate suppressed polylogarithmic factors.

\paragraph*{Previous Results}
One of the first results in this area is that of Clarkson, Eppstein, Miller, Sturtivant and Teng~\cite{DBLP:journals/ijcga/ClarksonEMST96}, 
who gave several randomized algorithms for computing approximate centerpoints,
under various conditions on the desired complexity and approximation quality. Chan~\cite{DBLP:conf/soda/Chan04} showed that the optimal centerpoint of an $n$-point set in $\R^d$ can be 
computed in time $O(n^{d-1})$. Miller and Sheehy~\cite{DBLP:journals/comgeo/MillerS10} gave a derandomized version of the Clarkson et al. algorithm, and subsequently their 
work was improved by Mulzer and Werner~\cite{DBLP:journals/dcg/MulzerW13}, who gave a slightly better quality bound on the obtained centerpoint. In 2019, Har-Peled 
and Jones~\cite{DBLP:journals/talg/Har-PeledJ21}, 
using a novel random walk based approach, 
improved the complexity of the Clarkson et al~\cite{DBLP:journals/ijcga/ClarksonEMST96} algorithm from $\tilde{O}(d^9)$ to $\tilde{O}(d^7)$,
where the $\tilde{O}$-notation indicates polylogarithmic factors in $d$. \\ 

In a remarkable breakthrough Cherapanamjeri~\cite{10756130} gave the first polynomial time algorithm  
for achieving $\Omega\pth{\frac{1}{d}}$-centerpoints, for a \emph{smoothed} version of the problem where the input points are perturbed by a small Gaussian noise. Further, 
for the original point set, the algorithm obtains a point which is at most $\e$-far from the halfspaces characterizing any $\Omega\pth{1/d}$-centerpoint. 
The algorithm is based on a gradient descent type approach, building on a connection between the Radial Isotropic Transformation~\cite{10756130} and testing an approximate centerpoint. 
While the degree of the polynomial running time is not explicitly stated in the paper, the algorithm (as described in ~\cite{10756130}, Fig. 1) runs for 
$T = 2^{80}(d/\e)^{16}$ steps, with each
step involving the computation of a Radial Isotropic Transformation which itself is a subroutine using e.g. the Ellipsoid method, and so takes more than unit time. 
The running time of the algorithm therefore seems to be prohibitively large 
for practical purposes, involving both a large multiplicative constant and a high degree polynomial in the ambient dimension. Thus, the question of finding an efficient and 
practical algorithm for computing $\omega\pth{1/d^2}$-centerpoints for arbitrary datasets, has remained open. \\

The existing algorithms for centerpoints are summarized in the table below.
These algorithms have found applications in related problems such as finding Tverberg points, center nets, 
etc.~\cite{DBLP:conf/esa/Har-PeledZ21,DBLP:journals/comgeo/MillerS10,DBLP:conf/swat/Har-PeledR24}.  \\

\begin{table}[h]
  \caption{Comparison of algorithms for finding a centerpoint}
  \centering
  \label{tab:center-algorithms}
  \begin{tabular}{|l|l|l|}
    \hline
    \textbf{Algorithm} & \textbf{Quality} & \textbf{Running time} \\ \hline
    Clarkson et al. (general) &
      $\bigl(\tfrac{1}{d+1}-\epsilon\bigr)$ &
      $O\!\bigl((\tfrac{d}{\epsilon^{2}}\log\tfrac{d}{\epsilon})^{d+O(1)}\log\tfrac{1}{\delta}\bigr)$ \\ \hline
    Clarkson et al. & & \\ 
    (polynomial in $d$) &
      $\displaystyle \frac{2/3e}{\bigl(d^{2}+3d+2\bigr)}$ &
      $\widetilde{O}(d^9)$ \\ \hline
    Chan (randomized; Las Vegas type) &
      $\displaystyle \frac{1}{d+1}$ &
      $O(n^{d-1})$ \\ \hline
    Miller \& Sheehy (deterministic) &
      $\displaystyle \frac{1}{2(d+1)^2}$ &
      $n^{O(\log d)}$ \\ \hline
    Mulzer \& Werner (deterministic) &
      $ \tfrac{1}{2(d+1)^2} $ &
      $f(2^{d+1}) + d^{O(1)} n$\footnotemark \\ \hline
    Mulzer \& Werner (deterministic) &
      $ \tfrac{1}{4(d+1)^3} $ &
      $O(d^{O(\log(d))}n)$ \\ \hline
    Har-Peled \& Jones &
      $\bigl(\tfrac{1-\vartheta}{2(d+2)^2}\bigr)$ &
      $\widetilde{O}(d^7)$ \\ \hline
    Cherapanamjeri & & \\ 
    ($(i)$ smoothed point-set or; \. & & \\ 
     $(ii)$ $\e$-far from all halfspaces & & \\ 
     characterizing centerpoints) &
      $\Omega\left(\frac{1}{d}\right)$ &
      $O(\mathrm{poly}(n, d, 1/\varepsilon, \log(1/\delta)))$ \\ \hline
    {\bf This paper}  &
      $\Omega\bigl(\tfrac{1}{d\log d\log\log d}\bigr)$ &
      $\tilde{O}_w(d^{5.5})$; \\ 
      & & $\tilde{O}(d^3e^{O\pth{\sqrt{d\log d}}})$. \\ \hline
  \end{tabular}
\end{table}
\footnotetext[1]{$f(m) = O(m^{d^2})$ is the time to compute a Tverberg point of depth $m/(d + 1)$ for $m$ points by
brute force.}

In spite of these different approaches and bounds, the best quality for centerpoints achievable with a running time polynomial in the ambient 
dimension, remains $O\pth{\frac{1}{d^2}}$. Currently the fastest algorithm achieving this quality is that of Har-Peled and Jones~\cite{DBLP:journals/talg/Har-PeledJ21}. While 
their output has nearly the same quality as that of Miller-Sheehy~\cite{DBLP:journals/comgeo/MillerS10} or Mulzer-Werner~\cite{DBLP:journals/dcg/MulzerW13}, their running time improves upon the previous 
results by a factor of $O(d^2)$. For obtaining $\omega\pth{\frac{1}{d^2}}$-centerpoints currently the fastest algorithm remains that of 
Clarkson et al.~\cite{DBLP:journals/ijcga/ClarksonEMST96}, who used linear programming to obtain $\pth{\frac{1}{d+1}-\e}$-centerpoints in time $O\pth{\pth{(d/\e)\log d/\e}^d}$.
It should be noted that since the algorithm uses linear programming, it is \emph{weakly polynomial}.  \\

\paragraph*{Centerpoint Algorithm of Har-Peled and Jones}
    Given a 
set of points in $\R^d$, a \emph{Radon partition} is a bi-partition of the set, such that the convex hulls of the two parts intersect and a \emph{Radon point} is a point in the intersection
of the two convex hulls. We now summarize some ideas used in centerpoint algorithms, especially the algorithm of Har-Peled and Jones. 

    Beginning with the seminal work of Clarkson et al.~\cite{DBLP:journals/ijcga/ClarksonEMST96}, several centerpoint finding algorithms have the following basic structure. 
Given an $n$-point set in $\R^d$, one 
maintains a set $Q$ of $n$ candidate points, which is initially the input set, and finally a set of $\alpha$-centerpoints, (where $\alpha\in (0,\frac{1}{d+1})$ depends  
on the specific algorithm). The goal is to modify the candidate set $Q$ so that for every halfspace $H$ of $\R^d$ that contains fewer than an $\alpha$-fraction of the points of $P$, no point 
of $Q$ lies in $H$ at the end of the procedure. 

Each step of the algorithm involves an \emph{exchange procedure}, where one or more current points are exchanged with new points computed during the step.
The procedure of Har-Peled and Jones~\cite{DBLP:journals/talg/Har-PeledJ21}, is particularly simple -- one chooses a random sample of $d+2$ points, 
and computes a Radon point of this sample.
Subsequently, the Radon point is added to the candidate set $Q$, and a previously independently chosen point is removed. The elegant analysis of the procedure is as follows.
Let $H$ be an $\alpha$-shallow halfspace, that is, a halfspace of $\R^d$ containing at most an $\alpha$-fraction of the input point set. With respect to membership in $H$, therefore, there are 
three possible outcomes of the exchange procedure: $(i)$ the deleted point was in $H$, and the new point was not in $H$, or $(ii)$ the deleted point was not in $H$, and the added point 
was in $H$, or $(iii)$ the deleted as well as the added points were both in $H$ or both outside $H$. The cardinality of $Q\cap H$, can therefore increase by one, decrease by one, or stay the 
same. As Har-Peled and Jones show, this can be modelled as a sticky random walk on the non-negative integers, which is strongly negatively biased, and therefore quickly converges to zero with
high probability. Taking a union bound over all the $O(n^d)$ equivalence classes of halfspaces (with respect to intersection with $P$), we get that with high probability, after a 
polynomial number of iterations, no point of $Q$ will remain in any such $\alpha$-shallow halfspace, and so with good probability, every point of $Q$ will be an $\alpha$-centerpoint.

The key observation that makes the above analysis work, is the fact that any halfspace containing a Radon point of a set $S$ of $d+2$ points, must also contain at least $2$ points of $S$. 
This means that for the random walk to go forward, at least two points lying in $H$ must have been chosen in the random sample. The probability of this happening (which is at 
most ${d\choose 2}\alpha^2$) is much less than that of the walk to go backwards, which requires that the point to be deleted was chosen from $H$ (which can happen with probability at most $\alpha$). 
This imbalance between the two probabilities results in the random walk being negatively biased and quickly converging to zero.

\subsection{Our Contribution}
We give the first weakly polynomial algorithm for computing $\omega\pth{\frac{1}{d^2}}$- centerpoints. Our centerpoints have quality $\Theta\pth{\frac{1}{d\log d\log\log d}}$, which 
improves on the long-standing $d^{O(d)}$ running time of Clarkson et al.~\cite{DBLP:journals/ijcga/ClarksonEMST96} for obtaining such centerpoints. Our algorithm also improves 
on the $\tilde{O}(d^7)$ running time of Har-Peled and Jones~\cite{DBLP:journals/talg/Har-PeledJ21}, although theirs is a strongly polynomial algorithm.

For low-dimensional point-sets, we also give an algorithm with complexity subexponential in the number of dimensions, and \emph{polynomial} in the size of the inputs.
This again improves on the bound of Clarkson et al.~\cite{DBLP:journals/ijcga/ClarksonEMST96} while not requiring inputs to have size polynomial in the number of points and dimensions.
Although this does not beat the complexity bound of Har-Peled and Jones~\cite{DBLP:journals/talg/Har-PeledJ21}, it is the first subexponential algorithm achieving a $\omega\pth{d^{-2}}$ quality 
of the centerpoint, and requiring time $o(d^d)$ in the ambient dimension. On the way, we also get a simple algorithm to compute $O\pth{\sqrt{\frac{n}{d}}}$-Tverberg 
partitions using a lifting trick of Sarkaria, together with sampling ideas from Sober\'on~\cite{DBLP:journals/combinatorica/Soberon19}. 
As in the earlier case, linear programming can also be used here to compute a point in the intersection of the parts contributing to the $s$-Tverberg partition. 

To state our results formally, let $P$ be a set of $n$ points in $\R^d$. 

\begin{theorem}
\label{thm:main-algo-bd}
    There exists a constant $c>0$ independent of $n$ and $d$, such that the following statements hold.
    \begin{itemize}
       \item There exists a randomized algorithm \Findcenter, such that given any $p_f\in (0,1)$, it computes a $\frac{c}{d\log d\log\log d}$-centerpoint of $P$ in weakly polynomial time 
        $O_w\pth{d^{5.5}\log^9 d\log(d/p_f)}$, and succeeds with probability at least $1-p_f$.
       \item There exists a randomized algorithm, which runs in time 
        \[O\pth{d^3\log^4 d\log(d/p_f)\cdot\pth{d^3\log^5 d+e^{O\pth{\sqrt{d\log d}}}}},\] 
        independent of the absolute values of its inputs, 
        and computes a $\frac{c}{d\log d\log\log d}$-centerpoint of $P$ with probability at least $1-p_f$.
    \end{itemize}
\end{theorem}

\paragraph*{Word RAM Model and Bit-Complexity Issues} At each iteration involving finding a Tverberg point, 
the output of the LP can have length polynomial in the lengths of the inputs. Since there are polynomially many steps (in $d$), and each iteration can 
use outputs of previous iterations, the final lengths of the bit-representations can become super-exponential in $d$. 
While this is not an issue in the Real RAM model, it appears in the Word RAM model, and can also be an issue in practical implementations. We give 
a simple Truncate-and-Proceed approach, which can effectively address this blow-up of bit complexities.
\begin{theorem}
\label{thm:trunc-proc}
   For any $p_f\in (0,1]$, there exists a polynomial function $f(d) = O(d^3\log^4d\log(d/p_f))$, 
such that given any $\e\in (0,1]$, while running \Findcenter, it suffices to truncate the output of each iteration of \TvbgPt after 
at most $O(f(d)+\log (1/\e))$ bits, so that the error in each point output by \Findcenter has $\ell_2$-norm at most $\e$. 
\end{theorem}

\paragraph*{Computing Tverberg Points} A related notion to the centerpoint of a point-set, is that of an \emph{$s$-Tverberg partition} of a point-set. 
Given a set $X$ of $r$ points in $\R^d$ and $s\geq 2 \in \mathbf{N}$, an \emph{$s$-Tverberg partition} of $X$ is a partition of $X$ into $s$ parts, such that the 
intersection of the convex hulls of the points in each part, is non-empty. An \emph{$s$-Tverberg point} of $X$, is a point in the common intersection of the convex hulls of 
the parts of an $s$-Tverberg partition of $X$. In the course of our analysis, we state and prove the following bound for computing $s$-Tverberg point for 
a set of $r$ points in $\R^d$, when $s = O\pth{\sqrt{r/d}}$.
        \begin{lemma}
        \label{l:s-tverb-partn-rand-constr}
        Given a set of $r$ points $X\subset \R^d$, with $r\geq \frac{16}{3}s^2d\log (3s)$, 
        a random $s$-coloring of $X$ is a Tverberg partition of $X$ into $s$ parts $\sqcup_{i=1}^s X_i$, with probability at least 
        $1-\exp\pth{-r/3s}$.
        Further, a point $c \in \bigcap_{i=1}^s X_i$ that lies in the common intersection of the convex hull 
        of the $s$ parts, can be computed in time $O_w\pth{r^{2.5}} $ with a weakly polynomial algorithm, or in time $O\pth{2^{O\pth{\sqrt{r\log r}}}+r^2(sd)}$ using an algorithm 
        independent of the absolute values of its input numbers.
        \end{lemma}

   The above lemma is useful when the number of points is not much larger than the dimension -- as for example in the case of high-dimensional datasets -- 
in which case it supersedes the bounds of Har-Peled and Zhou~\cite{DBLP:conf/esa/Har-PeledZ21} as well as previous bounds~\cite{DBLP:journals/comgeo/MillerS10,DBLP:journals/dcg/MulzerW13}, 
all of whose algorithms work when $s = \Omega\pth{r/(d+1)^2}$. In our 
case (see Section~\ref{sec:main-result-proof} for more details) we have $r = O\pth{d\log d\log \log d}$, and therefore $r/d^2 \ll 1$, 
so the previous algorithms would not work.

\paragraph*{Computing a colorful choice for the Colorful Carath\'eodory Theorem}
    Our algorithmic improvements in the Centerpoints and Tverberg problems, come from an improvement 
    of an important variant of the Carath\'eodory problem. The \emph{colorful} Carath\'eodory theorem, 
    given by Barany~\cite{BARANY1982141},  
    says that given a set of points partitioned into at least $d+1$ families, and a test point lying in their 
    convex hull, there exists a choice of a set having exactly one member of each family, such that the test point still lies in the 
    convex hull of the colorful choice. Both the existential and the algorithmic question of finding such a colorful 
    choice, have several applications~\cite{BaranyOnn97, BaranyOnn97_2}.  
    In particular, there have been a few approximate algorithms~\cite{mulzer_et_al:LIPIcs.SOCG.2015.44} proposed for the algorithmic question, and it continues to be an interesting open problem. It turns out our algorithm can be used to 
    improve the bounds of Mulzer and Stein~\cite{mulzer_et_al:LIPIcs.SOCG.2015.44}, for finding a colorful choice, in a wide range of the problem parameters.

\begin{theorem}[Colorful Carath\'eodory Algorithm]
    \label{thm:randomized_algo_colorful_carath}
    Let $F_1, \dots, F_N \subset \R^d$ with $0 \in \conv(F_i)$ and $|F_i| = O(d)$ for every
    $i$, and suppose a convex combination of the origin is given for each $F_i$. If
    $N \geq 6(d+1)d\log\!\big(6(d+1)^2\big) = \Theta(d^2\log d)$, then
    Algorithm~\ref{alg:random_colorful} outputs a Colorful Carath\'eodory transversal with
    failure probability $p_f \leq \exp\!\big(-N/(3(d+1))\big)$, in $O(N\,t_{\mathrm{Car}})$
    time, where $t_{\mathrm{Car}} = O(d^3 s + s^2) = O(d^4)$ is the cost of one Carath\'eodory
    reduction ($s = \max_i |F_i|$).
\end{theorem}

Finally, we give some applications of our results for algorithmic questions on centerpoints, such as the resilient consensus problem, finding subgradients of convex functions, etc.

\paragraph*{Overview of Our Algorithm}
   From the previous discussion on the centerpoint algorithm of Har-Peled and Jones, we observe that the key step utilizes the fact that a Radon point is a $2/(d+2)$-centerpoint
of a set of $d+2$ points, and the probability of the random walk going forward is essentially controlled by the upper bound on the probability that at least $2$ out of $d+2$ random  
points lie in a given halfspace. Now it may be asked, why not use a construction with a lower probability of the corresponding 
random walk drifting forward? For example could we compute a point such that any halfspace containing it must contain more points from our sample? 
This is exactly the idea that we implement.  \\ 

Thus, our overall algorithm structure follows that of ~\cite{DBLP:journals/talg/Har-PeledJ21}, except that at each step,
we shall compute the $s$-Tverberg point, ($s:= \log d$) of a random sample. Now for a fixed halfspace $H$, the random walk will go 
forward only if at least $s$ out of $d$ chosen points belong to $H$. 
While this seems like a simple plan -- and indeed a similar idea was used by Miller and Sheehy in their derandomization of the Clarkson et al. algorithm -- 
in practice there are several issues which need to be handled. The main issue is that it is not easy to compute an $s$-Tverberg point of a given point-set. Indeed, 
the problem of computing an $s$-Tverberg point is nearly equivalent to our original problem, since an $n/(d+1)$-Tverberg point is also a $1/(d+1)$-centerpoint. 
Consequently, computing Tverberg points is an area of active research~\cite{DBLP:conf/esa/Har-PeledZ21}. On the other hand, most results on Tverberg 
points focus on computing an $n/(d+1)$-Tverberg point 
whereas we only need a fast subroutine for the case $s:=\log d$, which does not seem to follow from existing algorithms.  

This motivates our second contribution -- a randomized algorithm which computes an $s$-Tverberg partition and the corresponding $s$-Tverberg point, for a given set 
of $\Omega(s^2d\log s)$ points in $\R^d$ (for arbitrary $s$), with high probability. The idea is to simply analyse a random partition and show that it is a Tverberg partition, 
via the lifting technique of Sarkaria~\cite{sarkaria1992tverberg}, using an idea implicit in Sober\'on~\cite{DBLP:journals/combinatorica/Soberon19}. Further, in the 
Sarkaria-Sober\'on framework, the problem of finding an $s$-Tverberg point,
reduces to finding the convex coefficients of a given point in the interior of a polytope with a linear number of vertices. This can be done in a standard way by solving a linear program.  

%
%
%

     \paragraph*{Outline of the paper} 
     In Section~\ref{sec:backgd} we provide some necessary background and preliminary details. In Section~\ref{sec:main-result-proof}, we give the main technical results and proofs. 
     Our improved algorithm for the Colorful Carath\'eodory problem, is stated and analysed in Section~\ref{sec:colorful_carath}.
     In Section~\ref{sec:gug} we give a generalized version of the Radon urn game and state a sufficient condition for it to succeed. 
     We also give applications of our results to some algorithmic questions on centerpoints, in Section~\ref{sec:appln}.

        \section{Background}
        \label{sec:backgd}

        We briefly introduce some of the definitions and tools needed for our results and proofs. 

        Given a set $X$ of points in $\R^d$, and all closed halfspaces in $\R^d$, the \emph{range space} $(X,\calR)$ denotes $X$ together with 
        the equivalence classes of closed halfspaces under the equivalence relation given by the subset of $X$ realised by intersection of the halfspace with $X$. 
        The equivalence classes are referred to as \emph{ranges}, and $\calR$ denotes their collection.
        For a range $R\in \calR$, and a subset $Y\subset X$, let $\mu_Y(R) := |R\cap Y|/|Y|$ denote the fraction of points of $Y$ in the halfspace $R$. 
        The \emph{VC dimension} of the range space $(X,\calR)$ is the maximum size of a subset $Y\subset X$, such that for all $Z\subset Y$, 
        $Z$ can be represented as the intersection of $Y$ with some range in $\calR$. The following fact is well-known about range spaces induced by points and halfspaces in $\R^d$: 
        the maximum number of distinct intersections of halfspaces with a set of $n$ points, is $O(n^{d})$ (see e.g.,~\cite{M99}[Chapter 5]). 

        For $\rho,\delta \in (0,1)$, a $(\rho,\delta)$-relative approximation for $X$ with respect to halfspaces, is a subset $S\subset X$, such that for 
        any range $R\in \calR$, 
        \[ (1-\delta)\mu_X(R) \leq \mu_S(R) \leq (1+\delta)\mu_X(R), \mbox{ if } \mu_X(R) \geq \rho \mbox{ and,}\]
        \[ |\mu_X(R)-\mu_S(R)|\leq \rho\delta, \mbox{ if } \mu_X(R) \leq \rho.\]
        The following bound of~\cite{DBLP:journals/jcss/LiLS01,DBLP:journals/dcg/Har-PeledS11} on the size of $(\rho,\delta)$-relative approximations will be useful for our analysis.
        \begin{lemma}
        \label{l:rel-approx-bd}
           Given a range space $(X,\calR)$ of VC dimension $d_0$, and $\rho,\delta,\tilde{p}_f\in (0,1)$, a random subset $S \subset X$ of  
         size $O\pth{\frac{1}{\delta^2\rho}\pth{\log \frac{d_0}{\rho}+\log\frac{1}{\tilde{p}_f}}}$ is a $(\rho,\delta)$-relative approximation with 
         probability at least $1-\tilde{p}_f$.
        \end{lemma}
        The next lemma gives a connection between a centerpoint for a range space $(X,\calR)$, and a centerpoint for its relative approximation $S$. 
        \begin{lemma}
        \label{l:cp-rel-approx}
        Given a set $X$ of $n$ points in $\R^d$ and a $(\rho,1/2)$-relative approximation $S$ of $X$, a $3\rho$-centerpoint for $S$ is a 
        $\rho$-centerpoint for $X$.
        \end{lemma}

        \begin{proof}[Proof of Lemma~\ref{l:rel-approx-bd}]
           Let $X$ and $S$ be a range space of points and halfspaces, and a $(\rho,1/2)$-relative approximation respectively, as stated in the Lemma.
         Let $c$ be a $3\rho$-centerpoint for $S$, and let $R$ be any closed halfspace containing $c$. By the definition of a centerpoint, $R\cap S$ has a $3\rho$-fraction of the points 
         of $S$, i.e., $\mu_S(R) \geq 3\rho$. Since $S$ is a $(\rho,1/2)$-relative approximation, this gives that $\mu_X(R)/2 \leq \mu_S(R)\leq (3/2)\mu_X(R)$, or 
         $\mu_X(R) \geq 2/3\mu_S(R) \geq 2\rho$. Further, for any range $R'$ with $\mu_X(R')\leq \rho$, we have that $\mu_S(R')\leq \rho+(1/2)\rho = (3/2)\rho$. Thus 
         $R'$ cannot contain $c$, since this would contradict that $c$ is a $3\rho$-centerpoint of $S$.
        \end{proof}

        \subsection{Tverberg Partitions and Sarkaria's Lifting Map}
        \label{subsec:tvb-part-lift-map}
         We are given a set $X$ of $r$ points in $\R^d$, and an integer parameter $s \geq 2$. For each $x\in X$, let $\tilde{x} := (x,1) \in \R^{d+1}$.
         Let $u_1,\ldots,u_s \in \R^{s-1}$ denote the vertices of the standard $s$-simplex in $\R^{s-1}$. That is, for each $i\in [s-1]$, $j\in [s-1]$, 
         $u_i(j) = 1$ if $j=i$, and $u_i(j) = 0$ otherwise. Lastly, $u_s= (-1,\ldots,-1)$. For each $i\in [r]$, define the family of points 
         $\calF_i := \{\tilde{x}_{i,j} := \tilde{x}_i \otimes u_j = \tilde{x}_iu_j^{\top}\;|\; j\in[s]\} \subset \R^{(s-1)(d+1)}$ \footnote{Here, the 
         matrix $\tilde{x}u_j^{\top}$ should be read as a $(d+1)(s-1)$-length column vector, by concatenating the columns of the matrix.}. Now, 
         let $\calF = \bigcup_i \calF_i$. Given a subset $T$ of points in the lifted set, $T \subset \cup_{i\in [r]} \calF_i$, we say 
         $T$ is an $s$-\emph{transversal} of $\calF$ if for every $i\in [r]$, $\exists  j: j \in [s]$ such that $\tilde{x}_{ij} \in T$. The following proposition, implicit in 
         Sarkaria~\cite{sarkaria1992tverberg} and Sober\'on~\cite{DBLP:journals/combinatorica/Soberon19}, gives the connection between $s$-part Tverberg partitions and transversals.

        \begin{lemma}
        \label{l:s-tverb-partn-sark-lift}
           If there exists an $s$-transversal $T = T(\calF)$ such that the origin $0 \in \conv(T)$, then $X$ has a Tverberg partition into $s$ parts. 
        \end{lemma}

        \begin{proof}[Proof of Lemma~\ref{l:s-tverb-partn-sark-lift}]
              The statement is well-known from the work of Sarkaria~\cite{sarkaria1992tverberg}, 
         see also Soberon~\cite{DBLP:journals/combinatorica/Soberon19}. Let $T = \{p_1,\ldots, p_r\}$ be a transversal of $\calF$ such that $0\in \conv(T)$. 
         Then there exist non-negative reals $\alpha_i$, $i=1,\ldots, r$, such that $\sum_i \alpha_i = 1$, and $\sum_i \alpha_i p_i = 0$. For each $j\in [s]$, let $S_j$ denote the 
         set of indices $i \in [r]$, such that $\tilde{x_i}\otimes u_j \in T$, for some $x_i \in X$. Now for each $j\in [s-1]$, consider the $j(d+1)$-th coordinate of the convex combination 
         $0 = \sum_i \alpha_i p_i$. By construction, this coordinate is $1$ for the points in $S_j$ and $-1$ for the points in $S_s$, otherwise it is zero. Thus, we get that for each $j\in [s-1]$, 
         \[ \sum_{i \in S_j} \alpha_i = \sum_{i\in S_s} \alpha_i.\]
         Since $\sum_{i\in [s]} \alpha_i = 1$, this means that for each $j\in [s]$, $\sum_{i\in S_j} \alpha_i = \frac{1}{s}$.
         Similarly if we consider for each $j\in [s-1]$, the weighted sum of the points of $T$ with coordinates restricted to $(j-1)(d+1)+1,\ldots, j(d+1)-1$, we get
         \[ \sum_{i\in S_j} \alpha_i x_i = \sum_{i\in S_s} \alpha_i x_i,\]
         where $x_i$ is the point of $X$ that was lifted to $p_i\in T$.
         Dividing the above equation by $\sum_{i\in S_j}\alpha_i$, i.e., by $1/s$, gives us that for every $j\in [s-1]$, there is a convex combination of subsets of the original points   
         $X_j := \{x_i\;:\; i \in S_j\}$, such that $\sum_{i\in S_j} \beta_i x_i = \sum_{i\in S_s} \beta_i x_i$, where $\beta_i = s\alpha_i$. (We check that for all 
         $j\in [s]$, $\sum_{i\in S_j} \beta_i = 1$, and for all $i\in [r]$, $\beta_i \geq 0$.) 
         This shows that 
         \[\bigcap_{j\in [s]} \conv(X_j) \neq \emptyset.\]
         Thus, $X = \sqcup_{j\in [s]} X_j$ is the desired partition of $X$ into $s$ parts, with non-empty intersection of their convex hulls. 
         \end{proof}

         \subsubsection{Computing an $s$-Tverberg partition}
         \label{subsubsec:comput-s-tvb-part}
         From the above randomized construction, we see that if the transversal $T = T(\calF)$ is such that $0\in \conv(T)$, then we obtain a partition of $X$ into $s$ parts $X_1,\ldots,X_s$, 
         whose convex hulls intersect. Further, given a convex combination $(\alpha_i)_{i\in [r]}$ (with $\alpha_i\geq 0$ and $\sum_{i\in [r]}\alpha_i =1$), 
         such that $\sum_{i\in [r]} \alpha_i p_i = 0$, for each $j\in [s]$, the point $c(T):=\sum_{i\in S_j} \alpha_i x_i$ lies in $\bigcap_{j\in [s]} X_j$.  \\
         
         Therefore, given a transversal $T$ such that $0\in \conv(T)$, we need to find the convex combination $(\alpha_i)_{i\in [r]}$, such that 
         $\sum_i \alpha_i p_i = 0$, where $\alpha_i$'s are non-negative reals whose sum is $1$. In order to accomplish this, the problem can be 
         formulated as a \emph{linear programme} (LP). 
         In general, a linear programme is an optimization problem over real-valued variables, having a set of 
         linear constraints and a linear cost function over the variables, which is to be optimized. It is a major open problem in the theory of algorithms, to find whether there exists a strongly 
         polynomial algorithm for solving linear programme. The current fastest algorithms for linear programmes over $n$ variables, with a total size $l$ of the input 
         in bits, run in weakly-polynomial time $O_w(n^{2.5}l)$ or even less 
         (under commonly-held complexity-theoretic assumptions~\cite{63499,DBLP:conf/colt/LeeSZ19,DBLP:conf/stoc/Jiang0WZ21}). A subexponential algorithm independent of the size of 
         the inputs is also known~\cite{10.1145/142675.142678}; for $r$ variables and $D$ constraints, the algorithm takes time $O\pth{r^2D+e^{O\pth{\sqrt{r\log r}}}}$.
         
         The LP formulation of the problem of finding a suitable convex combination $(\alpha_i)$, is given below.
 
         \begin{eqnarray} 
             \mbox{maximize } 1, \mbox{ such that} \label{eqn:opt-func} \\
             \forall i,\; \alpha_i &\geq& 0, \label{eqn:cond1}\\ 
             \sum_i \alpha_i &=&1,           \label{eqn:cond2}\\
             \sum_i \alpha_i p_i &=& 0.      \label{eqn:cond3}
         \end{eqnarray} 
         The above LP either finds a convex combination $(\alpha_i)$ satisfying the constraints, or returns $FAIL$ if no such combination exists.
        

	\section{Main Result}
	\label{sec:main-result-proof}

In this section we describe and analyse our algorithmic routines for finding centerpoints. \\

\begin{algorithm}
\caption{Compute $\alpha$-centerpoint of given point set $P$ (\Findcenter)}
\label{Algo:alpha-centerpoint}
\begin{algorithmic}[1] 
\Require Point set $P \subset \R^d$, $p_f\in (0,1]$, $\alpha<1/\tilde{O}(d)$.
\Ensure $Q=Q(P)$, i.e., a set of $\alpha$-centerpoints of $P$.
\State Initialize $Q:= P$, $s:=\log d$
\For{$i=1$ to $\tilde{c}|P|\log |P|\cdot d\log(d/p_f)$}
    \State Pick a set $S$ of $t= Cs^2d\log(sd)$ points randomly from $Q$
    \State Compute $p_S :=\TvbgPt(S,s)$
    \State Delete a random point from $Q$
    \State Add $p_S$ to $Q$
\EndFor
\Return $Q$
\end{algorithmic}
\end{algorithm}

\begin{algorithm}
\caption{Compute a Tverberg $s$-partition of $S$ (\TvbgPt)}
\label{Algo:s-Tverberg}
\begin{algorithmic}[1]
\Require  Point set $S \subset \R^d$, $s \in \N$.
\Ensure $p_S$, a point in the intersection of the convex hulls of an $s$-partition of $S$.
\State Apply Sarkaria's lift to each point $p\in S$ to get the family $\calF_i :\{\tilde{x}_{i,j}\;:\; j\in [s]\}$
\State Construct random transversal $T = \{\tilde{t}_1,\ldots,\tilde{t}_{|S|}\}$
\For{$i=1$ to $|S|$}
    \State Choose $\tilde{t}_i$ uniformly randomly from $\calF_i$
\EndFor
\State Find convex combination $(\lambda_i)$ such that $\sum_{i\in [N]}\lambda_i \tilde{t}_{i} = 0$, using the LP given by ~\eqref{eqn:opt-func}-~\eqref{eqn:cond3}
\If{LP returns FAIL, i.e., $\not\exists$ $(\lambda_i)_{i\in [|S|]}$ such that $\sum_{i\in [N]}\lambda_i \tilde{t}_i = 0$}
    \State Return FAIL
\EndIf
\Return $p_S = s\cdot\pth{\sum_{i=1}^{|S|} \tilde{t}_i \cdot [\tilde{x}_{i,1} == \tilde{t}_i]}$\footnotemark
\end{algorithmic}
\end{algorithm}

\footnotetext{Here we use the Iverson bracket notation, i.e., $[a] = 1$ if the 
        condition $a$ holds, and $[a] = 0$ otherwise.}
\vspace{2mm} 

Let us begin with the analysis of Algorithm \rm{TverbergPoint}, which is also the constructive proof of Lemma~\ref{l:s-tverb-partn-rand-constr}.
         \begin{proof}[Proof of Lemma~\ref{l:s-tverb-partn-rand-constr}]
         The construction is implicit in Sober\'on~\cite{DBLP:journals/combinatorica/Soberon19}. Here we make precise the lower bound on the required number of points $r$, in terms of 
         the number of parts $s$, and the ambient dimension $d$. We first apply the Sarkaria lift to $X$, to get a family $\calF= \bigcup_{i\in [r]}\calF_i \subset \R^{(s-1)(d+1)}$. 
         Next, we construct a random transversal $T$ of $\calF$ by picking one point from each family $\calF_i \in \calF$. 
         We need to check that the transversal $T$ contains the origin in its convex hull. Let $\calH_0 = \calH_0(\calF)$ be the set of equivalence classes of closed halfspaces 
         in $\R^{(s-1)(d+1)}$ containing the origin, distinguished by their intersection with points in $\calF$. 
         The following condition is necessary and sufficient for the origin to lie in the convex hull of $T$.
         \begin{proposition}
         $0\in \conv(\calF)$, if and only if for every closed halfspace $H\in \calH_0$, $T\cap H \neq \emptyset$. 
         \end{proposition}

         Let us fix a 
         halfspace $H\in \calH_0$, and consider $\calF_i \in \calF$ for some $i\in [r]$. Since the family $\calF_i$ contains the origin in its convex 
         hull (as $\sum_{j\in [s]} \frac{1}{s}\tilde{x}_{ij} = 0$), therefore there exists $j\in [s]$, such that $\tilde{x}_{ij} \in H$. Since with probability 
         $1/s$, $\tilde{x}_{ij}\in T$, therefore we get that with probability at least $1-1/s$, $T\cap H\cap \calF_i \neq \emptyset$. As this holds for each 
         $i\in [r]$ independently, we get that 
         \[ \Prob{T\cap H = \emptyset} \leq \pth{1-\frac{1}{s}}^{r} \leq \exp\pth{-r/s} .\]
         From the bound on the number of distinct intersections of a point-set with halfspaces in $\R^d$ (see Section~\ref{sec:backgd}), it follows that 
         \[|\calH_0| \leq 2{|\calF|\choose (s-1)(d+1)} \leq 2{rs\choose (s-1)(d+1)} \leq \pth{\frac{2er}{d+1}}^{(s-1)(d+1)}\]
         for $s\geq 2$.
         Now taking the union bound over all $H\in\calH_0$, we get that 
         \begin{eqnarray}
            \Prob{\exists H \in \calH_0\; : \; T\cap H = \emptyset} &\leq& \exp\pth{-r/s}\cdot \pth{\frac{2er}{d+1}}^{(s-1)(d+1)} \\
                                                                    &\leq& \exp\pth{-r/s+2sd\log(r/d)}. 
         \end{eqnarray}
         Substituting $r\geq \frac{16}{3}s^2d\log s$, we check that the above probability is at most $\exp\pth{-r/3s}$. Thus, with probability at least $1-\exp\pth{-r/3s}$, 
         every halfspace containing the origin has a non-empty intersection with $T$, and therefore $0\in \conv(T)$. 

         Finally, a point lying in the intersection of the $s$ parts can be found by solving the linear programme described in Section~\ref{subsubsec:comput-s-tvb-part}.
         From well-known bounds on solving linear programmes~\cite{63499,DBLP:conf/colt/LeeSZ19}, this can be done via a weakly polynomial algorithm that takes $O_w(r^{2.5})$, 
         Another option is to use a subexponential time algorithm~\cite{10.1145/142675.142678} that is independent of the size of the input numbers in bits. From the 
         bound in Section~\ref{subsubsec:comput-s-tvb-part}, using $r$ variables and $D:= (s-1)(d+1)\approx sd$ constraints, we get that the subexponential algorithm takes 
         time $O\pth{r^2ds+\exp\pth{O\pth{\sqrt{r\log r}}}}$.
        \end{proof}

The next theorem gives the correctness and running time analysis of the \Findcenter algorithm. The theorem, together with the proof of Lemma~\ref{l:s-tverb-partn-rand-constr}, will allow us 
to prove our main theorem, Theorem~\ref{thm:main-algo-bd}. 
\begin{theorem}[Analysis of \Findcenter]
\label{thm:center-pt-algo-analysis}
There exists $\tilde{c}>0$ such that for any $p_f\in (0,1]$, with probability at least $1-p_f$, after $\tilde{c} n\log n\cdot (d\log (n/p_f))$ iterations of 
Algorithm \Findcenter, $n= O\pth{d^2\log d\log (d/p_f)\log\log d}$, all points of $Q$ are 
$\alpha$-centerpoints, where $\alpha := c'(d\log d\log\log d)^{-1}$, and $c'>0$ is independent of $n$ and $d$. 
\end{theorem}

Given the above theorem, the proof of Theorem~\ref{thm:main-algo-bd} can be completed as below.
\begin{proof}[Proof of Theorem~\ref{thm:main-algo-bd}]
   Let $\rho := (d\log d\log\log d)^{-1}$, and $p_f\in (0,1]$ be as stated in the theorem. By Lemma~\ref{l:rel-approx-bd}, a random subset $S$ of $P$, of size 
$\Omega\pth{\rho^{-1}d\log (2\rho^{-1}/p_f)} = \Omega\pth{d^2\log d\pth{\log (d/p_f)}\log\log d}$ is a $(\rho,1/2)$-relative approximation with probability at least $p_f/2$.
Now we apply Theorem~\ref{thm:center-pt-algo-analysis}, with failure probability $p_f/2$, to get a total failure probability (failure during the $(\rho,1/2)$-relative approximation 
construction, or during the call to Theorem~\ref{thm:center-pt-algo-analysis}) of $p_f$.
The running time of the algorithm is determined by the $(i)$ number of iterations of the \rm{For} loop, which is at most $\tilde{c}dn\log n \log (24n/p_f)$, where $n= O(d^2\log d\log(d/p_f)\log\log d)$, 
and $(ii)$ the time taken for \rm{TverbergPoint} to execute with $r:= t = Cs^2d\log s$, each time it is called. 
Using the fast weakly-polynomial algorithm for linear programmes, this takes time $O_w(r^{2.5}) = 
O_w(d^{2.5}\log^{5} d(\log\log d)^{2.5})$. 
Summarizing, we get that the running time complexity can be bounded as  
\begin{eqnarray}
   & &      \#(\mbox{iterations of } {\rm For} \mbox{ loop}) \notag \\ 
   & &      \times \;\; (\mbox{Maximum time taken per iteration of {\rm TverbergPoint}}) \notag \\
   &\leq &  O\pth{d(d^2\log d\log (d/p_f)\log\log d)\log(d^2\log d\log (d/p_f)\log\log d)\log(24(d^2\log d\log(d/p_f)\log\log d/p_f)} \notag \\
   & &      \times \;\; O_w\pth{d^{2.5}\log^5d(\log\log d)^{2.5}} \notag \\
   &= &     O\pth{d^3\log^3d\log(d/p_f)\log\log d}\times O_w\pth{d^{2.5}\log^5d(\log\log d)^{2.5}} \label{eqn:num-iter} \\
   &= &     O_w\pth{d^{5.5}\log^8d\log(d/p_f)(\log\log d)^{3.5}}. \notag
\end{eqnarray}

In order to avoid issues with the bit-size complexity of the inputs, we can also use the algorithm of ~\cite{10.1145/142675.142678}, which is subexponential 
in $d$ but independent of the sizes of the inputs, 
to compute the $s$-Tverberg point during the calls to $\rm{TverbergPoint}$. Similar to the previous analysis, this has a running time of  
\begin{eqnarray}
   & &      \#(\mbox{iterations of } {\rm For} \mbox{ loop}) \notag \\ 
   & &      \times \;\; (\mbox{Maximum time taken per iteration of {\rm TverbergPoint}}) \notag \\
   &= & O\pth{d^3\log^4 d\log (d/p_f)} \times O\pth{r^2ds + \exp\pth{O\pth{\sqrt{r\log r}}}} \label{eqn:use-num-iter}\\
   &= & O\pth{d^3\log^4 d\log (d/p_f)} \\ 
   & & \times O\pth{d^2\log^4d\cdot d\log d(\log\log d)^2 + \exp\pth{O\pth{\sqrt{d\log^2 d\log d(\log\log d)}}}} \label{eqn:subs-rs-val}\\
   &= & O\pth{d^3\log^4 d\log (d/p_f)\pth{d^3\log^5d(\log\log d)^2 + \exp\pth{O\pth{\sqrt{d\log^3 d\log\log d}}}}}, \notag 
\end{eqnarray}

where in ~\eqref{eqn:use-num-iter}, we used the expression for the number of iterations derived 
previously in ~\eqref{eqn:num-iter}, and in ~\eqref{eqn:subs-rs-val}, we substituted $r=d\log^2 d\log\log d$ and $s = \log d$.
By Lemma~\ref{l:cp-rel-approx}, $c$ is also an $\Omega\pth{\frac{1}{d\log^2 d\log\log d}}$-centerpoint for $P$.
\end{proof}

\subsection{Analysis of \Findcenter}

    Consider the equivalence classes generated by halfspaces in $\R^d$, under the equivalence condition given by intersection with $P$. 
Let $\calH= \calH(P)$ denote the set of equivalence classes of halfspaces, whose intersection with $P$ is at most $\alpha n$.
Fix $H\in \calH$. We shall show that with probability at least $1-p_f/|\calH|$, after $O(n\log^2 n)$ iterations, no point of $Q$ remains in $H$.
\begin{lemma}
   \label{l:fixed-halfsp-probab}
   There exists $\tilde{c}>0$ such that, given any $p_f\in (0,1]$ and $n= O\pth{d^2\log^2 d\log(d/p_f)\log\log d}$, 
   with probability at least $1-p_f/2$, after at most $\tilde{c}dn\log n \log (24n/p_f)$ iterations of the \rm{For} loop in Algorithm \Findcenter, $H\cap Q = \emptyset$.
\end{lemma}

Assuming Lemma~\ref{l:fixed-halfsp-probab}, Theorem~\ref{thm:center-pt-algo-analysis} follows easily. Applying the union bound over all halfspaces $H\in \calH$, 
the probability that there exists a halfspace $H\in \calH$, such that after $\Omega\pth{n\log n\log (24n|\calH|/p_f}$ iterations, $Q\cap H \neq \emptyset$, is at most 
$|\calH|\cdot \frac{p_f/2}{|\calH|} = p_f/2$. Therefore, with probability at least $1-p_f/2$, $Q\subset \bigcap_{H\in\calH} \bar{H}$.
That is, any halfspace containing a point of $Q$ must have more than $\alpha n$ points of $P$. Thus, each point of $Q$ will be an 
$\alpha$-centerpoint of $P$.

To complete the proof of the theorem therefore, it only remains to prove Lemma~\ref{l:fixed-halfsp-probab}.

\begin{proof}[Proof of Lemma~\ref{l:fixed-halfsp-probab}]
In order for the steps of \Findcenter to succeed, it is sufficient for the following events to occur:
\begin{itemize}
    \item Event $T_1$: Every call to \rm{TverbergPoint} successfully computes an $s$-Tverberg point.
    \item Event $T_2$: After $\tilde{c}dn\log n \log (24n/p_f)$ iterations of the \rm{For} loop in \Findcenter, we have $H\cap Q = \emptyset$.
\end{itemize}

Thus, the probability of failure can be bounded by $\Prob{\overline{T_1\cap T_2}}$, which is at most $\Prob{\overline{T_1}} + \Prob{\overline{T_2}\cap T_1}$.
That is to say, the algorithm \Findcenter fails if $(i)$ there exists an iteration where \rm{TverbergPoint} fails to compute an $s$-Tverberg point of its 
input, or $(ii)$ after $\tilde{c}dn\log n\log (24n/p_f)$ iterations of the \rm{For} loop in \Findcenter, we still do not have $H\cap Q=\emptyset$, even though 
\rm{TverbergPoint} has run correctly each time. We shall show that both these probabilities, i.e., $\Prob{T_1^c}$ and $\Prob{T_1\cap T_2^c}$ are at most $p_f/4$. 
Thus, we shall have $\Prob{T_1\cap T_2} \geq 1-p_f/2$, as required.\\

We first bound $\Prob{\overline{T_1}}$. This can be done easily by taking the failure probability from Lemma~\ref{l:s-tverb-partn-rand-constr}, and applying 
a union bound over the number of iterations. In each iteration, we call \rm{TverbergPoint} with $r := t = Cs^2d\log s \leq C_1d\log^2 d\log\log d$ points, for $s:=\log d$ 
and taking $C_1=2C$. By Lemma~\ref{l:s-tverb-partn-rand-constr}, the probability of failure is at most $\exp\pth{-r/3s} = \exp\pth{-C_2d\log d\log\log d}$, taking $C_2= C_1/3$.
Using that $n= O\pth{d^2\log^2 d\log(d/p_f)}$, the number of iterations is at most 
\begin{eqnarray}
   \tilde{c}dn\log n\log (24n/p_f) &=& \tilde{c}_1d^3\log^2 d\log (d/p_f)\cdot \log d\cdot \log (24d^2\log^2 d\log(d/p_f)/p_f) \notag \\ 
                                  &=& \tilde{c}_2d^3\log^3 d\log^2 (d/p_f).
\end{eqnarray}
The probability that some iteration failed, is therefore at most 
\[ \Prob{T_1^c} \leq \tilde{c}_2d^2\log^3 d\log^2(d/p_f)\cdot\exp\pth{-C_2d\log^2 d} \leq \exp\pth{-C_3d\log^2 d+ 2\log\log(d/p_f)},\]
where we take $C_3 = C_2/2$ to accommodate the multiplicative terms outside the exponent, except for the factor $\log^2(d/p_f)$, which appears directly in the exponent 
as $2\log\log(d/p_f)$. Now using the condition $d\log^2 d \gg \log\log d/p_f$, we get that for an appropriately large $C$ (or sufficiently large $d$), 
\[\Prob{T_1^c} \leq \exp\pth{-C_3d\log^2 d+2\log\log (d/p_f)} \ll p_f/4.\]

We now arrive at the estimation of the probability $\Prob{T_2^c\cap T_1}$, i.e., the \rm{For} loop fails, even though \rm{TverbergPoint} succeeded each time.  
The action of the algorithm on the points of $Q\cap H$ can be understood as follows. Initially, we have $Q=P$, so a random point of $Q$ lies in $Q \cap H$ 
with probability at most $\beta_0:=\alpha$. In each iteration, we pick a point $q\in Q$ at random, and mark $q$ for deletion. 
Then we choose a set $S$ of $t$ random points with replacement from $Q$, compute an $s$-Tverberg point $p$ of $S$, and replace $q$ from $Q$ with $p$. 
Let $\alpha_i$ denote the fraction of points of $Q$ in $H$ in the beginning of the $i$-th iteration. 
If $p\in H$ and $q\not\in H$, then $|Q \cap H|$ increases after the iteration, so the fraction $\alpha_{i}$ goes up. Otherwise, if $p\not\in H$ and $q\in H$, $\alpha_i$ decreases. 
In all other cases, $\alpha_i$ stays unchanged. \\
Since $p$ has Tverberg depth $s$, it can lie in $H$ only if at least $s$ of the chosen $t$ points are in $H$.
Thus, the probability of the fraction $\alpha_i$ of points of $Q$ in $H$ going up in the $i$-th iteration, is upper bounded by the probability that 
$|S\cap H| \geq s$.
This suggests the following \emph{$s$-Tverberg urn game} could be used to analyse the evolution of the distribution of the number of points in $Q\cap H$.

\paragraph*{$s$-Tverberg Urn Game}
The set up is as follows. We have an urn $U$ with $b$ red balls and $n-b$ black balls. In each iteration, we perform the following experiment.
\begin{enumerate}
  \item Choose a random ball from $U$, mark it for deletion, and return it to $U$. 
  \item Select a set $S$ of $t$ balls at random, chosen with replacement from $U$.
  \item If at least $s$ of the chosen $t$ balls are red, insert a red ball back into the urn
  \item Else, insert a black ball into $U$.
  \item End if.
  \item Return the chosen $t$ balls to $U$.
  \item Remove the marked ball from $U$.
\end{enumerate}

The connection between Algorithm \Findcenter and the $s$-Tverberg urn game 
is based on the observation that 
for any $i\in \N$, the number of points in $Q\cap H$ after $i$ iterations of Algorithm \Findcenter is stochastically dominated by the 
number of red balls $b_i$ in the urn $U$, after $i$ iterations of the urn game. More precisely, we have the following.

\begin{proposition}
\label{propn:algo-vs-urn-iter}
For any $i>0$, the distribution of the number of points of $Q\cap H$ after $i$ iterations is stochastically dominated by the distribution of the number of red balls in the urn 
after $i$ iterations of the above urn game. 
\end{proposition}

\begin{proof}[Proof of Proposition~\ref{propn:algo-vs-urn-iter}]
   The formal proof follows by induction on $i$. To start with, we have $|Q\cap H| = \beta_0n = \alpha n = b$, so the fraction of points of $Q$ in $H$, is equal to the fraction of red balls 
in $U$. Assume this holds for $i$ iterations.  
Let $b_i$ denote the number of red balls in the urn $U$ after $i$ iterations of the game have been played. 
In Algorithm \Findcenter, let $S_i$ denote the random sample chosen in iteration $i$, and let $p_i$ denote the $s$-Tverberg point of $S_i$ computed by the algorithm.
Let $q_i$ denote the point chosen for deletion from $Q$, in the $i$-th iteration.
In the $(i+1)$-st iteration, we have that $\beta_i$ increases in iteration $i$, if $p_i \in H$, and $q\not\in H$.
The event $p_i \in H$ can happen only if at least $s$ points of $S_i$ were in $H$. Thus the probability that $\beta_{i+1}\geq \beta_i$ is at most the probability that 
$s$ red balls were chosen from $U$ in iteration $i$ in the $s$-Tverberg urn game. \\

%
\end{proof}

Observe that the probability that at least $s$ balls out of $t$ chosen ones are red, can be bounded as follows.
\begin{claim}[Probability of having $\geq s$ out of $t$ balls red]
\label{clm:un-bd-s-outof-t}
   Given a coin with probability $p\in [0,1]$ of falling Heads, the probability that there are at least $s$ Heads out of $t$ independent coin tosses, 
is at most ${t\choose s}p^s$.
\end{claim}

\begin{proof}
   The required event occurs if and only if \emph{there exists} a set $Y \subset [t]$, $|Y|=s$, such that all the coin tosses with indices in $Y$, come Heads.
By the union bound over all subsets of $[t]$ of size $s$, this probability is at most ${t\choose s}$ times the probability that a fixed $s$-sized subset 
comes out all Heads, which is $p^s$. The claimed bound follows.

\end{proof}

Since the probability of a single ball being red is simply the fraction of red balls in the urn, i.e., $\beta_i$, 
therefore by Claim~\ref{clm:un-bd-s-outof-t}, the probability $q(\beta_i) \leq {t\choose s}\beta_i^s$. Thus, the $s$-Tverberg urn game is 
an instantiation of the Generalized urn game with $q(\beta_i) = {t\choose s}\beta_i^s$. 
Thus we get that the condition to apply Theorem~\ref{thm:gen-urn-game-summary}, 
is satisfied if $q(\beta_i) \leq {t\choose s}\beta_i^s \leq \frac{\beta_i}{2-\beta_i} \leq \beta_i$ (here the last inequality holds since $\beta_i\leq 1$). In other
words, $\pth{\frac{et\beta_i}{s}}^s\leq \beta_i$, or $\beta_i\leq \pth{\frac{s}{et}}^{s/(s-1)}$. Since $\beta_i$ is the fraction of red balls out of $n$ balls in the 
urn, we get that if the $s$-Tverberg game starts with at most $n\cdot\frac{s}{3et}\cdot\pth{\frac{s}{et}}^{1/(s-1)} 
= n\cdot\frac{O(1)}{d\log d\log\log d} = \alpha n$ balls, then the probability 
that after $\tilde{c}n\log n\log(24n/p_f)$ iterations, there are no red balls left in the urn is at least $1-p_f/4$. By Proposition~\ref{propn:algo-vs-urn-iter}, this is 
at least the probability that $Q\cap H=\emptyset$ after $\tilde{c}n\log n\log(24n/p_f)$ iterations of Algorithm \Findcenter. Thus the probability 
$\Prob{T_1\cap T_2^c}$ is at most $p_f/4$, which completes the proof of Lemma~\ref{l:fixed-halfsp-probab}.
\end{proof}

\subsection{Addressing Bit-Complexity via Truncate-and-Proceed}
In this section we discuss the Truncate-and-Proceed approach to bounding the bit-size complexity and give the proof of Theorem~\ref{thm:trunc-proc}. 
The idea is that truncating an input after $l$ bits, introduces a maximum error of $2^{-l}$. Since LP algorithms typically operate with a given error 
parameter $\e$, we shall show that given any such $\e>0$, it is possible to truncate the output at each stage, 
after polynomially many bits in $d$ and $\log(1/\e)$, such that the error in the final output, is at most $\e$. 

After each iteration $i$, we truncate the output after $\log (1/\e')$ bits.

For $i=1,2,\ldots k$, with $k= O\pth{d^3\log^4 d\log(d/p_f)}$ denoting the number of iterations of \TvbgPt, 
let $\e_i$ denote the maximum error after $i$ iterations of \TvbgPt. Let $\tilde{\lambda}_{ij}$, $j=1,\ldots,r$ be the convex 
combination output at the $i$-th iteration. We have $\tilde{\lambda}_{ij} = \lambda_{ij}+\eta_{ij}$, where for $j=1,\ldots, r$, 
$\lambda_{ij}$ is the exact solution, and $\eta_{ij}$ is the error introduced by the LP algorithm. We shall choose the error parameter in the LP algorithm 
such that for each $j$, $|\eta_{ij}| \leq \e'$, where $\e' = O\pth{(d\log^2d)^{-k}\e}$.
Similarly, for $j\in [r]$, let $p^*_{ij} = p^{ij}+\zeta_{ij}$ be the points used as input in the $i$-th iteration of \TvbgPt. Starting with $\e_0 = 0$, 
we define the parameter $\e_i$ to be such that $\|\zeta_{ij}\|_2 \leq \e_i$.

The error introduced in the output of the $i+1$-st iteration of \TvbgPt at most the error induced due to the previous errors, together with  
the error due to truncation of the output. This can be bounded as below.
\begin{eqnarray*}
   & & \|\sum_j \tilde{\lambda}_{ij}p^*_{ij} - \sum_j \lambda_{ij}p_{ij}\|_2 + 2^{\log (1/\e')} \\
   &= & \|\sum_j \eta_{ij}p_{ij} + \sum_j \zeta_{ij}\tilde{\lambda}_{ij}\|_2 + \e' \\
   &\leq & \|\sum_j \eta_{ij}p_{ij}\|_2 + \|\sum_j \zeta_{ij}\tilde{\lambda}_{ij}\|_2 + \e', 
\end{eqnarray*}
where in the last step we used the triangle inequality. The first term is at most 
\[ \max_j |\eta_{ij}|\cdot\|\sum_j p_{ij}\|_2  \leq \e'\cdot r\sqrt{d},\]
where we assumed the original points had norm $\leq 1$, so that the points $p_{ij}$ after lifting, had norm at most $\sqrt{d}$.  
The second term is at most 
\[ \max_j \|\zeta_{ij}\|\cdot \max_j |\tilde{\lambda}_{ij}| \leq 2\e_i. \]

Thus the total error in the output at the $i+1$-st iteration, increases as 
\[ \e_{i+1} \leq 2\e_i + r\sqrt{d}\e' + \e' = (3+r\sqrt{d})\e'.\]
This gives that after $k= O\pth{d^3\log^4d\log(d/p_f)}$ iterations, the final error is at most 
\[ \e_k \leq (3+r\sqrt{d})^k\e'.\]

Thus, choosing $\e' = O(r^{-k}) = O\pth{(d\log^2d)^{O(d^3\log^4d\log(d/p_f))}}\e$, suffices 
to bound the cumulative error in the final output, by $\e$.

\section{Computing a Colorful Choice for the Colorful Carath\'edory Problem}
\label{sec:colorful_carath}

In this section, we shall see how our algorithmic ideas can be applied to the setting of 
the Colorful Carath\'eodory Theorem. The ``colorful" version of Carath\'eodory's theorem,  
can be stated as follows.

\begin{theorem}[Colorful Carath\'eodory Theorem]
    \label{thm:colorful_caratheodory}
    Let $\mathcal{F} = \{F_1, \dots, F_N\}$ be a family of sets with $F_i \subset \R^d$
    for every $i$ and $N \geq d+1$. If $0 \in \conv(F_i)$ for all $F_i \in \mathcal{F}$,
    then there exists a transversal $T$, consisting of exactly one point $f_i \in F_i$
    from each set in $\mathcal{F}$, such that $0 \in \conv(T)$.
\end{theorem}
Several algorithms already exist for computing a Colorful Carath\'eodory transversal.
Mulzer and Stein~\cite{mulzer_et_al:LIPIcs.SOCG.2015.44} gave one that runs in $d^{O(\log d)}$ time when the
family is large enough.
\begin{theorem}[Mulzer--Stein~\cite{mulzer_et_al:LIPIcs.SOCG.2015.44}, Theorem~4.4]
    \label{thm:mulzer_stein}
    Let $F_1, \dots, F_N \subset \R^d$ be $N = \Theta(d^2 \log d)$ sets such that each $F_i$
    embraces the origin (i.e.\ $0 \in \conv(F_i)$) and $|F_i| = O(d)$. Then, given a convex
    combination of the origin for each $F_i$, one can compute a colorful Carath\'eodory
    transversal $T$ in $d^{O(\log d)}$ time.
\end{theorem}
We show that in the same regime a randomized algorithm matches the family-size
requirement of Theorem~\ref{thm:mulzer_stein} in polynomial time. The random-choice argument
is due to Sober\'on~\cite{SOBERON_2018} (Lemma~2.2), who states it as an existence result for the
colorful Carath\'eodory theorem via a union bound over the halfspaces through the origin. We
phrase it as a Monte Carlo algorithm with an explicit failure probability and running time.
\begin{lemma}
    \label{lem:random_choice}
    Let $N = |\mathcal{F}|$ and $s = \max_{F_i \in \mathcal{F}} |F_i| \geq 2$. A random
    colorful transversal $T$ (one point drawn uniformly and independently from each $F_i$)
    is a Colorful Carath\'eodory choice with probability at least
    \[
        1 - \exp\!\pth{-\frac{N}{s} + 2d\log\!\pth{\frac{Ns}{d}}}.
    \]
\end{lemma}
\begin{proof}
    Construct $T$ by choosing one point uniformly and independently at random from each
    $F_i$, and consider the probability that $0 \notin \conv(T)$. Observe that
    $0 \notin \conv(T)$ exactly when there is a closed halfspace $H$ containing the origin
    such that $T \cap H = \emptyset$; since $0 \in \conv(F_i)$, every such $H$ contains at
    least one point of each $F_i$ (otherwise $\conv(F_i)$ would lie strictly on the other
    side of $H$, excluding the origin).

    Let $P = \bigcup_i F_i$, so $|P| \leq Ns$. With respect to membership it suffices to
    consider the equivalence classes of halfspaces induced by their intersection with $P$,
    of which there are at most
    \[
        2\binom{Ns}{d} \;\leq\; 2\pth{\frac{eNs}{d}}^{d}.
    \]
    Fix one such halfspace $H$. Each of the $N$ independent draws lands outside $H$ with
    probability at most $1 - 1/|F_i| \leq 1 - 1/s$, so
    \[
        \Prob{T \cap H = \emptyset} \;\leq\; \pth{1 - \frac{1}{s}}^{N}
        \;\leq\; \exp\!\pth{-N/s}.
    \]
    Taking a union bound over all the $O((Ns)^d)$ equivalence classes of halfspaces (with
    respect to intersection with $P$),
    \begin{align*}
        \Prob{\exists\, H : T \cap H = \emptyset}
            &\leq \exp\!\pth{-N/s} \cdot 2\pth{\frac{eNs}{d}}^{d} \\[2pt]
            &= \exp\!\pth{-\frac{N}{s} + d\log\!\pth{\frac{Ns}{d}} + d + \log 2} \\[2pt]
            &\leq \exp\!\pth{-\frac{N}{s} + 2d\log\!\pth{\frac{Ns}{d}}},
    \end{align*}
    where the last step uses $d\log(Ns/d) \geq d + \log 2$, valid whenever
    $Ns/d \geq e^{\,1 + (\log 2)/d}$ and in particular throughout our regime. Taking the
    complement proves the claim.
\end{proof}
The algorithm reduces each class to at most $d+1$ points and then samples one point from
each. The reduction is the constructive form of Carath\'eodory's theorem: given the
coefficients of a convex combination of the origin, Mulzer and Stein's
Lemma~2.1~\cite{mulzer_et_al:LIPIcs.SOCG.2015.44} returns a $0$-embracing affinely independent subset
$F_i' \subseteq F_i$ in $O(d^3|F_i| + |F_i|^2)$ time, using only Gaussian elimination and no
linear program.
\begin{algorithm}[H]
\caption{Randomized Colorful Carath\'eodory Choice (Monte Carlo)}
\label{alg:random_colorful}
\begin{algorithmic}[1]
    \Require $\mathcal{F} = \{F_1, \dots, F_N\}$ with $0 \in \conv(F_i)$ for all $i$, a convex
             combination of the origin for each $F_i$, and $N \geq 6(d+1)d\log(6(d+1)^2)$
    \Ensure A colorful transversal $T$ with $0 \in \conv(T)$ (with high probability)
    \For{$i = 1$ to $N$}
        \State Reduce $F_i$ to $F_i' \subseteq F_i$ with $|F_i'| \leq d+1$ and
               $0 \in \conv(F_i')$ (Lemma~2.1~\cite{mulzer_et_al:LIPIcs.SOCG.2015.44})
        \State Pick a point $f_i \in F_i'$ uniformly at random
    \EndFor
    \Return $T = \{f_1, \dots, f_N\}$
\end{algorithmic}
\end{algorithm}
We are now ready to prove Theorem~\ref{thm:randomized_algo_colorful_carath}.
\begin{proof}[Proof of Theorem~\ref{thm:randomized_algo_colorful_carath}]
    After reduction, $\max_i |F_i'| \leq d+1$. Applying Lemma~\ref{lem:random_choice} with
    $s = d+1$, it suffices that $-N/(d+1) + 2d\log(N(d+1)/d) \leq -N/(3(d+1))$, i.e.\
    $\log(N(d+1)/d) \leq N/(3(d+1)d)$; this holds for $N \geq 6(d+1)d\log(6(d+1)^2)$ since
    then $N/(3(d+1)d) \geq 2\log(6(d+1)^2) = \log((6(d+1)^2)^2)$, while
    $\log(N(d+1)/d) = \log(6(d+1)^2\log(6(d+1)^2)) \leq \log((6(d+1)^2)^2)$. The running time
    is $N$ reductions of cost $t_{\mathrm{Car}}$ each, the random draw being negligible.
\end{proof}
Thus the algorithm matches the $\Theta(d^2\log d)$ family size of Mulzer and Stein under the
same $O(d)$ class-size assumption, but runs in polynomial time rather than $d^{O(\log d)}$.
If zero error is required, one may verify $0 \in \conv(T)$ by a single linear-program
feasibility test and repeat until success, giving a Las Vegas algorithm.

        \section{Generalized Urn Game}
        \label{sec:gug}

In this section we describe and analyse the generalized urn game that is the heart of the analysis of our algorithms.
\paragraph*{Generalized Urn Game $\calG$}
The set up is as follows. Initially we have an urn $U$ with $I_0/2$ red balls and $n-I_0/2$ black balls, where $I_0\leq n/2$. We then perform a number of iterations of the following experiment.
Let $\beta_i$ denote the fraction of balls in $U$ having red color in the beginning of the $i$-th iteration.
We are given a function $q=q(\beta_i)$.
The $i$-th iteration proceeds as follows.
\begin{enumerate}
  \item Choose a random ball from the urn $U$, mark it for deletion, and return it to $U$. 
  \item Toss a coin with probability of Heads at most $q=q(\beta_i)$. 
  \item If it falls Heads, insert a red ball into $U$.
  \item Else, insert a black ball into $U$.
  \item End if.
  \item Remove the marked ball from $U$.
\end{enumerate}
The following theorem on the generalized urn game $\calG$ shall be crucial in the analysis of our algorithm. 
\begin{theorem}
\label{thm:gen-urn-game-summary}
   Let $I_0$ be the maximum $i$ such that $\pth{q(\beta_i) \leq \frac{\beta_i }{2-\beta_i}}$. Suppose $I_0 \leq n/2$.
   Then there exists $\tilde{c}>0$ such that, given any $p_f\in (0,1]$, 
   with probability at least $1-p_f$, the urn $U$ starting with $I_0/2$ red balls and $n-I_0/2$ black balls, has only black balls left 
   after $\tilde{c}n\log n \log (6n/p_f)$ iterations of $\calG$.
\end{theorem}

\begin{proof}[Proof of Theorem~\ref{thm:gen-urn-game-summary}]
   After each iteration of the generalized urn game $\calG$, the number of red balls in the urn $U$, can 
   increase, decrease, or stay unchanged. Let us count only the steps where the number of balls in the 
   urn changes, by the sequence of random variables $(Z)_i$, $i\in \N$. If the ball being removed from $U$ has the same color 
   as the ball being added, $Z_i$ does not change and we do not increase $i$. Otherwise we increase $i$, and 
   if a red ball is being removed from the urn, and a black ball is being added, we have $Z_{i+1} = Z_i+1$, and  
   if a black ball is being added and a red ball is being removed, we have $Z_{i+1}=Z_i-1$.
   In one iteration, the probability that the number of red balls increases is $q^+(i) := q(\beta_i)(1-\beta_i)$, and that it decreases  
   is $q^-(i):= (1-q(\beta_i))\beta_i$.
   Thus, conditioned on the number of red balls not staying the same in an iteration, the behaviour of the $(Z)_i$'s can be modelled by the following random walk. 

   \[ Z_{i+1} = \left\{\begin{array}{ll}
                        Z_i + 1 & \mbox{ with probability } q^+(i)/(q^+(i)+q^-(i)) \\
                        Z_i - 1 & \mbox{ with probability } q^-(i)/(q^+(i)+q^-(i)) \\
                       \end{array}\right. \]
   The random walk $Z_i$ starts at $Z_0=I_0/2$, and ends either at $M = I_0$, in which case it is said to have \emph{failed}, or at zero, in which 
   case it has \emph{succeeded}. An easy observation from the assumption of the Theorem, gives us the following lemma.
   \begin{lemma}
   \label{l:qbi-qpl-qm}
      If $q(\beta_i) \leq \frac{\beta_i}{2-\beta_i}$, then $q^+(i) \leq q^-(i)/2$.
   \end{lemma}
   \begin{proof}[Proof of Lemma~\ref{l:qbi-qpl-qm}]
      Substituting the condition in the definitions of $q^+(i)$ and $q^-(i)$, we get 
   \[q^+(i) = q(\beta_i)(1-\beta_i) \leq \frac{\beta_i(1-\beta_i)}{2-\beta_i}, \mbox{ and} \] 
   \[q^-(i) = (1-q(\beta_i))\beta_i \geq \frac{2\beta_i(1-\beta_i)}{2-\beta_i},\] 
   from which the claimed inequality follows.
   \end{proof}
   
   Therefore, we get that under the condition of Lemma~\ref{l:qbi-qpl-qm},  
   \[ Z_{i+1} = \left\{\begin{array}{ll}
                        Z_i + 1 & \mbox{ with probability } \leq 1/3 \\
                        Z_i - 1 & \mbox{ with probability } \geq 2/3. 
                       \end{array}\right. \]
   Next, we bound the number of times the random walk visits any particular value.
   The following standard probabilistic lemmas are adapted from the proof of Har-Peled and Jones. 
   \begin{lemma}
   \label{l:lem-3-pt-2-hpj}
      Given $p_f\in (0,1]$ and any integer $j\in [0,I_0]$, the probability that the random walk $(Z)_i$ visits $j$ more than $10\log_2 (10/p_f)$ times, is at most $p_f$.
   \end{lemma}

   Using the above lemma, we can bound the probability that the random walk ever fails. 
   \begin{lemma}
   \label{l:lem-3-pt-3-hpj}
      Given $p_f\in (0,1]$, if $I_0 \geq 20\log_2 (10/p_f)$, then the probability that $(Z)_i$ ever visits $I_0$, is at most $p_f$.
   \end{lemma}
   By an argument similar to that of the well-known \emph{coupon collector's problem}, we can bound the number of times an iteration of 
   $\calG$ does not change the number of red balls in the urn $U$.
   \begin{lemma}
   \label{l:lem-3-pt-4-hpj}
      Given $p_f\in (0,1]$ and any $j\leq I_0$, there are at most $f(\beta_j) := O\pth{(1/\beta_j)\log (20/p_f)}$ iterations of $\calG$, such that the number of 
      red balls in $U$, remains $j$. 
   \end{lemma}
   Finally, we can bound the total number of iterations of the urn game $\calG$, and the probability that it succeeds.
   \begin{lemma}
   \label{l:lem-3-pt-6-hpj}
      Given $p_f\in (0,1]$, with probability at least $1-p_f$, the generalized urn game $\calG$ beginning with $I_0/2$ red balls and $n-I_0/2$ black balls in the urn $U$, 
    has at most $O\pth{(n\log n)\log (1/p_f)}$ iterations, after which no red balls are left in $U$. 
   \end{lemma}

\end{proof}

\section{Applications}
In this section we give some applications of our centerpoint algorithm.


\label{sec:appln}

\subsection{Centerpoints for the Resilient Vector Consensus Problem}
\label{subsec:centerpoints}

The \emph{consensus problem} asks a set of distributed processes to agree on a single output value even when up to $X$ of them behave adversarially.
When each process starts with a vector in $\mathbb{R}^d$ for $d \ge 2$, the task is called the \emph{resilient vector consensus} (RVC) problem.
A solution to RVC is correct if every non-faulty process eventually decides on the same vector and that vector lies within the convex hull of the non-faulty processes' initial vectors.

\paragraph*{Background}
Abbas, Shabbir, Li and Koutsoukos~\cite{ABBAS2022110046} introduced a centerpoint-based aggregation rule that strictly improves the resilience guarantees of several earlier RVC algorithms. 
We restate the relevant definition and theorem of Abbas, Shabbir, Li and Koutsoukos and show how our algorithm strengthens their \begin{proposition}[{\cite[Proposition 4.10]{ABBAS2022110046}}]
\label{prop:abbas-prop}
Given a set of $N_i$ points in general positions in $\R^d$, of which $F_i < F_{max}$ are adversarial, then it is possible to compute an $F_i$-safe point using centerpoint if
\begin{equation*}
    F_{max} = 
    \begin{cases}
        \lfloor \frac{N_i}{d+1} \rfloor, & \text{for } d=2,3, \text{ and} \\
        \Omega\pth{\frac{N_i}{d^2}}, & \text{for } d>3.
    \end{cases}
\end{equation*}
\end{proposition} 
Our attention centers on \emph{approximate distributed robust convergence} (ADRC), a technique for achieving resilient vector consensus whose fault-tolerance is 
determined by the maximum number of adversarial neighbours each process may have. By relaxing this neighbourhood bound, our centerpoint routine raises ADRC's 
resilience threshold further. 

\paragraph*{Network model.}
The communication structure is a directed graph
$G=(V,E)$ whose vertices are the processes.  An edge
$(i,j)\!\in\!E$ means that process~$j$ can read the current state (or
\emph{vector}) of process~$i$.  The \emph{neighbourhood} of~$j$ is
$P_j=\{i : (i,j)\in E\}$, of size $N_j=\lvert P_j\rvert$;

\begin{definition}[Safe point]
\label{def:safe}
Given a set $P \subset \mathbb{R}^d$ with at most $X$ adversarial points, a point 
$s \in \mathbb{R}^d$ is called \emph{$X$-safe} if $s$ is guaranteed to lie in the interior of the convex hull of the $(|P|-X)$ non-adversarial points.
\end{definition}

\begin{theorem}
\label{thm:centerpoint-to-safe}
Let $P\subset\mathbb{R}^d$ with $|P|=N$.  
If $N > X/\alpha$, then every $\alpha$-centerpoint of $P$ is $X$-safe, and every $X$-safe point is an $\alpha$-centerpoint with $\alpha = X/N$.
\end{theorem}

\begin{proof}
We prove the forward direction; the converse is immediate from the definitions.

Assume $c$ is an $\alpha$-centerpoint of $P$.  
By definition, every closed halfspace containing $c$ also contains at least $\alpha N$ points of~$P$. Note that if any halfspace contains $\alpha N$ points contains more than $X$ points as $\alpha N > \alpha * X/\alpha = X$. It means that if we have X adversarial points(vectors) and in every halfspace that contains c we have more than X points it means that each halfspace containing c contains at least one vector from a non-adversarial process which means that it lies in the convex hull of every non adversarial process's vector.
Thus $c$ is $X$-safe.
\end{proof}

\begin{theorem}
\label{thm:local-neighbourhood}
For process~$i$, suppose its neighbourhood $P_i$ in $\mathbb{R}^d$ satisfies  
\[
X_i < \alpha N_i .
\]
Then any $\alpha$-centerpoint of $P_i$ is an $X_i$-safe point (a vector) for a process~$i$.
\end{theorem}

\begin{corollary}
\label{cor:new-bound}
Applying our centerpoint algorithm, which achieves  
$\alpha = \mathcal{O}\!\left(\frac{1}{d\log d\log\log d}\right)$,
Theorem~\ref{thm:local-neighbourhood} guarantees safety provided
$X_i \;<\; \mathcal{O}\!\Bigl(\frac{N_i}{d\log d\log\log d}\Bigr)$.
Thus the tolerable number of adversarial processes in each neighbourhood is 
$\mathcal{O}\!\left(\frac{N_i}{d\log d\log\log d}\right)$.
\end{corollary}

The next couple of applications are improvements of bounds from Har-Peled and Jones~\cite{DBLP:journals/talg/Har-PeledJ21}, by applying our centerpoint algorithm.
\subsection{Algorithm for lowerbounding a function with a gradient oracle}

We now show how a centerpoint algorithm, paired with a gradient oracle, can produce a lower bound for a convex function $f$.

\begin{definition}
The subdifferential of a convex function $ f : \mathbb{R}^d \to \mathbb{R} $ at a point $ x \in \mathbb{R}^d$ is the set of all vectors $ g \in \mathbb{R}^d $ (called subgradients) such that:
\[f(y) \geq f(x) + \langle g \,, y-x\rangle, \quad \forall y \in \mathbb{R}^d.\]

The subdifferential at $ x $, is denoted by $ \partial f(x) $
\end{definition}

\begin{theorem}
Let $f : \mathbb{R}^d \to \mathbb{R}$ be a convex function, which is not necessarily differentiable. Consider a set $X \subset \mathbb{R}^d$, $|X| = n$. Assume the existence of 
$(i)$, an oracle that for any point $x \in \mathbb{R}^d$, returns a random element from the subdifferential $\partial f(x)$, and 
$(ii)$, an algorithm capable of computing a centerpoint of quality $O(\alpha)$, where $\alpha$ is a given parameter.
Then, using $O(\alpha^{-1} \log(n))$ queries to the oracle, it is possible to compute a point $z \in \mathbb{R}^d$ such that 
$f(z) \leq f(x), \quad \forall x \in X$.
\end{theorem}

\begin{proof}

The proof follows an iterative reduction scheme similar to binary search. Initialize $ X_1 = X $, the original point set. For each iteration $ i \geq 1 $
\begin{enumerate}
\item Compute an \( O(\alpha) \)-centerpoint \( c_i \) of \( X_i \).

\item Query the subgradient oracle to obtain \( g_i \in \partial f(c_i) \). Now look at all the points that satisfy $f(x) < f(c_i)$, $x\in \mathbb{R}^d$. Notice that 
$0 > f(x) - f(c_i) \geq \langle g \,, x-c_i,\rangle$
which is true from the subgradient definition. We can now see that all the points that satisfy $f(x) < f(c_i)$ satisfy $\langle g_i, x - c_i \rangle < 0$.

\item Let $H_{c_i}$ denote the open halfspace $ \{ x \in \mathbb{R}^d \mid \langle s_i, x - c_i \rangle < 0 \} $, which contains $\{ x \in \mathbb{R}^d \mid f(x) < f(c_i) \}$. Let $H_{c_i}^c$ denote the complement of $H_{c_i}$ If $|X_i \cap H_{c_i}^c| \geq \alpha*|X_i|$ which holds if $c_i$ is a correct $\alpha$-centerpoint as $c_i \in H_{c_i}^c$, we will delete all of the points in this halfspace (we are discarding points for which $f(x) \geq f(c_i)$) and get\[X_{i+1} =  X_i \cap H_{c_i}.\]Otherwise we will repeat the i-th iteration as the centerpoint the algorithm calculated was not correct. 

\end{enumerate}
The process terminates when \( |X_i| \leq O(1) \). The algorithm then returns $m$ such that 
\[
f(m) = \min\left\{ \min_{1 \leq j \leq i} f(c_j), \, \min_{x \in X_i} f(x) \right\}.
\]
Since $f$ is convex, it follows that $f(m) \leq f(x),  \forall x\in X$.\\  
We can now show that the above algorithm will end. We know that the expected number of unsuccessful iterations before we succeed is constant. 
We also know that in each successful iteration we will be removing an $\alpha$ fraction of points. This implies that the number of iterations $\theta$ satisfies 
$ (1-\alpha)^\theta |X| \leq O(1)$ so that $\theta = O(\alpha^{-1} \log(|X|))$.
\end{proof}

\begin{corollary}
Assuming we use our algorithm we could substitute
    $\alpha = O\pth{\dfrac{1}{d\log d\log\log d}}$
and would need $O(d\log d\log\log d\log n)$ queries to the oracle.
\end{corollary}

\subsection{Algorithm for deciding whether a convex body is $\e-heavy$ or $\e-light$}
Assume we have access to an oracle for a given convex body $B$ . When queried with a point $p$, the oracle behaves as follows:

\begin{itemize}
  \item If $p \in B$, the oracle confirms that $p \in B$.

  \item If $p \notin B$, the oracle returns a separating hyperplane $h$. One of the halfspaces created by the hyperplane contains $B$ and the other contains the point $p$. \par
\end{itemize}

\begin{theorem}
Let $P$ be a set of $n$ points in $\mathbb{R}^d$, and let $\varepsilon > 0$ and $\lambda > 0$ be parameters and $\alpha$ be the quality of a centerpoint algorithm. 
Choose a random sample $S \subseteq P$ of size
\begin{equation}
    s =O\Bigl(\e^{-1}\bigl(d \alpha^{-1} \log \e^{-1} \log(\alpha^{-1} \log\e^{-1}))\log\e^{-1}+\log \lambda^{-1}\bigr)\Bigr) =  \tilde{O}\Bigl(d \alpha^{-1} \e^{-1}\Bigr)
\end{equation}
There is an iterative algorithm which, given oracle access to a convex body $B$ can determine whether it is $\e$-light or $\e$-heavy to $P$ 
($\lvert B\cap P\rvert < \varepsilon\,n$ or $\lvert B\cap P\rvert \geq \varepsilon\,n$) using $S$ and $O(\alpha^{-1} \log(\e)^{-1})\alpha^{-1}\e)$ queries to the oracle.
\end{theorem}
\begin{corollary}

Substituting $\alpha = O\left(\frac{1}{d \cdot \log d\log\log d}\right)$ from our centerpoint algorithm, we obtain the condition:
$s = \tilde{O}\Bigl(d^3 \log (d \e^{-1})\Bigr)$
and the number of queries to be $O(d \cdot \log d\log\log d \log \e^{-1})$. 
\end{corollary}
We will prove the theorem later on. Now we will describe how the algorithm would work. Let $S_0 = S$
\begin{enumerate}
  \item
        Compute an $\alpha$-centerpoint $c_{i}$ of $S_{i-1}$.

  \item 
        Ask the oracle whether $c_{i} \in B$.
        \begin{itemize}
          \item If $c_{i}\in B$, $c_{i}$ return that $B$ is heavy.
          \item Otherwise the oracle supplies a separating hyperplane $h_{i}$.  
                Let $h^{-}_{i}$ be the open halfspace that contains $c_{i}$.
        \end{itemize}

  \item
        Remove that halfspace: $S^{temp}_{i} \;=\; S_{\,i-1}\setminus h^{o}_{i}$.

  \item 
        If the $\alpha$-centerpoint returned by our algorithm is a valid $\alpha$-centerpoint, than $|S^{temp}_{i}|\le (1-\alpha)\,|S_{\,i-1}|$ and we set $S_{i} =  S^{temp}_{i}$;  
        otherwise we keep the old set, $S_{i} =  S_{\,i-1}$ and try again assuming the error probability is constant.
We stop when $S_i \leq \e|S|$.
To prove our theorem we will first have to prove a couple of lemmas.
\end{enumerate}
\begin{lemma}
    The expected number of iterations will be $O(\alpha^{-1} \log(\e)^{-1})$.
\end{lemma}
\begin{proof}
    In each iteration where the algorithm returns a valid $\alpha$-centerpoint which happens with constant probability. We 
delete an $\alpha$-fraction of the set $S_i$. It means that the expected number of iterations is $x$, where $(1-\alpha)^x \leq \e$, so $x = O(\alpha^{-1} \log(\e)^{-1})$. 
\end{proof}

\paragraph{Statement on AI usage} AI was not used in obtaining the results of this article or writing them up. For the full version, ChatGPT was used in checking for typographical errors.


\bibliography{bibliography}
\appendix 
\paragraph*{Appendix} 
   Here we give the proofs of the lemmas occurring in the proof of Theorem~\ref{thm:gen-urn-game-summary}.
   \begin{proof}[Proof of Lemma~\ref{l:lem-3-pt-2-hpj}]
      Let $t_f:= \arg\min_{t}\{Z_t= j\}$. Since the probability of the random walk $(Z)_i$ advancing forward is $q^+(i)\leq 1/3$, and of going backward is $q^-(i)\geq (2/3)$ for any 
    $i\in \N$, we have that for a fixed $k\in \N$, 
      \[ \Prob{Z_{t_f+2k}=j|Z_{t_f}=j} \leq {2k\choose k}(1/3)^k(2/3)^k \leq (8/9)^k.\]
    Therefore, summing over all $k\geq k_0$, the probability $p_{k_0}$ that the random walk ever visits $j$ after the first $2k_0$ steps after the first visit, 
   can be bounded as 
   \[ p_{k_0} \leq \sum_{k\geq k_0} (8/9)^{k} = 9(8/9)^{k_0}.\]
   Now setting $k_0 = 10\log_2(10/p_f)$, we get that 
   \begin{eqnarray}
     p_{k_0} &\leq& 9\cdot 2^{-\log_2(9/8)\cdot 10\log_2(10/p_f)} \;\;<\;\; p_f, \label{eqn:pk0-bd}
   \end{eqnarray}
   which is the claim of the lemma.
   \end{proof}
   \begin{proof}[Proof of Lemma~\ref{l:lem-3-pt-3-hpj}]
      The proof follows from the proof of Lemma~\ref{l:lem-3-pt-2-hpj}, by observing that the random walk $(Z)_i$ starts at $Z_0= I_0/2$, and so it can arrive at $I_0$
   only at time $t\geq I_0/2$. Therefore from~\eqref{eqn:pk0-bd} in the proof of Lemma~\ref{l:lem-3-pt-2-hpj}, if $(I_0/2)\geq 10\log(10/p_f)$, then the probability $p_{I_0}$ that the 
   random walk ever visits $I_0$, given that it started at $I_0/2$, is at most $p_f$.
   \end{proof}

   \begin{proof}[Proof of Lemma~\ref{l:lem-3-pt-4-hpj}]
      Under the premise of Theorem~\ref{thm:gen-urn-game-summary}, conditioned on the game not failing, we always have $\beta_i\leq 1/2$.
   Therefore we get that for all $i$ in the duration of the game, 
   \[ q^+(i)+q^-(i) = (1-q(\beta_i))\beta_i + q(\beta_i)(1-\beta_i) = q(\beta_i)(1-2\beta_i)+\beta_i \geq \beta_i,\]
   since $1-2\beta_i\geq 0$. This gives that in expectation, the number of steps of $\calG$ before the ratio $\beta_i$ changes,  
   is $1/(q^+(i)+q^-(i)) \leq 1/\beta_i$.
   By Markov's inequality, the probability that there are at most twice this number of iterations of $\calG$ before $\beta_i$ changes, is at most $1/2$. 
   Now we can have random variables $X_1,\ldots, X_k$, which indicate the number of iterations the game $\calG$ spends each time it visits $i$. 
   By Lemma~\ref{l:lem-3-pt-2-hpj}, $k\leq 10\log_2(20/p_f)$, with probability at least $1-p_f/2$. Using a standard Chernoff bound for the 
   sum of independent geometric random variables (see e.g.,~\cite{DBLP:journals/talg/Har-PeledJ21} (Corollary 3.5)), we get that with probability at least $1-p_f/2$, 
   the sum $\sum_{j=1}^k X_j$ is at most $4$ times its expected value, i.e., $40(1/\beta_j)\log(20/p_f)$, with probability at least $1-p_f/2$. 
   Taking the union bound over the failure probabilities from the application of Lemma~\ref{l:lem-3-pt-2-hpj} and the Chernoff bound, we get the claimed bound 
   in the lemma.
   \end{proof}

   \begin{proof}[Proof of Lemma~\ref{l:lem-3-pt-6-hpj}]
      The proof follows from $(i)$ the proof of Lemma~\ref{l:lem-3-pt-4-hpj}, by summing the time bound from the lemma over all $i< I_0$, and taking the failure probability to be 
   at most $p_f/2$, $(ii)$ taking the failure probability in Lemma~\ref{l:lem-3-pt-3-hpj} to be also $p_f/2$, and $(iii)$, taking union bound over the probability of the bad events 
   that the random walk ever fails, and that it takes time more than $O\pth{n\log n\cdot\log (1/p_f)}$, given that it does not fail. 
   \end{proof}

\end{document}